\documentclass[a4paper,twoside,10pt]{letter}
\usepackage{graphicx,saj,multicol,subeqnarray}

\def\papertype{}

\def\udc{...}
\begin{document}
\baselineskip=3.1truemm
\columnsep=.5truecm
\newenvironment{lefteqnarray}{\arraycolsep=0pt\begin{eqnarray}}
{\end{eqnarray}\protect\aftergroup\ignorespaces}
\newenvironment{lefteqnarray*}{\arraycolsep=0pt\begin{eqnarray*}}
{\end{eqnarray*}\protect\aftergroup\ignorespaces}
\newenvironment{leftsubeqnarray}{\arraycolsep=0pt\begin{subeqnarray}}
{\end{subeqnarray}\protect\aftergroup\ignorespaces}
%

% Running titles

\markboth{\eightrm PROPAGATION OF HIGH FREQUENCY WAVES}
{\eightrm A. ANDI{\' C}}

{\ }

\publ

\type

{\ }

% Title

\title{PROPAGATION OF HIGH FREQUENCY WAVES IN THE QUIET SOLAR ATMOSPHERE}

% Authors

\authors{A. Andi{\' c}$^{1,2}$}

\vskip3mm

% Address

\address{$^1$HiROS,School of Physics and Astronomy, College of Engineering and Physical Sciences, The University of Birmingham, Edgbaston, Birmingham, B15~2TT, UK}

\address{$^2$ Astrophysics Research Centre, School of Mathematics and Physics, Queen's University, Belfast, BT7~1NN, UK}

% Received and Accepted dates

\dates{August 22, 2008}{October 6, 2008}

% Abstract

\summary{High-frequency waves ($5$mHz to $20$mHz) have previously been suggested as a source of energy accounting partial heating of the quiet solar atmosphere. The dynamics of previously detected high-frequency waves is analysed here. Image sequences are taken using the German Vacuum Tower Telescope (VTT), Observatorio del Teide, Izana, Tenerife, with a Fabry-Perot spectrometer. The data were speckle reduced and analysed with wavelets. Wavelet phase-difference analysis is performed to determine whether the waves propagate. We observe the propagation of waves in the frequency range $10$mHz to $13$mHz. We also observe propagation of low-frequency waves in the ranges where they are thought to be evanescent in regions where magnetic structures are present.}

% Keywords (see keywords.pdf file)

\keywords{SUN:oscillations -- SUN:atmosphere}

\begin{multicols}{2}
{

% Sections

\section{1. INTRODUCTION}

An ongoing debate about the heating mechanism of the solar atmosphere and the role of waves  in it (Ghosh, 2002) present the motivation for studying the dynamics of high-frequency waves. This work is a continuation of the wave analysis, parts of which are already presented in previous papers (Andic, 2007a, 2007b and Andjic, 2006) and focuses on waves with frequencies from $1$mHz to $22$mHz. Special attention will be given to interpretation of the results for the frequencies above $10$mHz.\par

The temperature of the solar atmosphere varies from a minimum value in the photosphere (around $4 \cdot 10^3$K) to a maximum typically found in the corona ($1 \cdot 10^6$K).  What supplies the energy necessary to form this temperature difference is still under discussion. Some authors claim that high-frequency waves carry the necessary energy (Ulmschneider, 1971a, 1971b, 2003; Stein and Leibacher, 1974; Kalkofen, 1990, 2001; Wedemeyer-B\"{o}hm et al. 2007) while other authors claim that low-frequency ones are the main source (Wang, Urlich and Coroniti, 1995; Jefferies et al. 2006).\par

 Recent work by Fossum and Carlsson (2005) and Andic (2007a) found no observational evidence for flux energetic enough to promote the acoustic heating proposed for high-frequency waves (Ulmschneider, 1971a, 1971b, 2003; Stein and Leibacher, 1974; Kalkofen, 1990, 2001). Following these observational results Wedemeyer-B\"{o}hm et al. (2007) present calculations using 3D models (Wedemeyer et al., 2004) and state that high-frequency acoustic waves do in fact have a role in the energy supply to the corona. \par

A simple assumption is that acoustic waves will propagate upward, form into a shock and therefore dissipate energy. This explanation assumes that the magnetic field is not required for the propagation of acoustic waves. Work by Rosental et al. (2002) and De Pontieu et al. (2004) have shown otherwise. Rosental et al. (2002) present 2D model of wave propagation in the  presence of a magnetic field, and they conclude that the presence of magnetic fields significantly complicates the waves and their associated dynamic. Jefferies  et al. (2006) states that waves which were previously considered to be  evanescent (Thio, 2006) can propagate when magnetic fields are present.

\section{2. OBSERVATIONS}

The data presented here are the same as already presented in previous papers (Andic, 2007a, 2007b; Andjic, 2006). \par

The spectral lines Fe I $543.45$nm ($g_{L}=0$) and $543.29$nm ($g_{L}=0.335$) were used and 2D spectroscopy was performed using the German Vacuum Tower telescope (VTT), Observatorio del Teide, Izana, Tenerife, with the Fabry-Perot spectrometer. The data were obtained during the mornings of 22 and 24 June 2004, with excellent seeing conditions and with the use of an  adaptive optics system (Berkefeld, Soltau and von der Luehe, 2003). The solar disk centre was targeted and the data were obtained in bursts of images. The exposure time of an individual image was $30$ms and the cadence between successive images was $0.25$s. The time taken to scan the line profile was $28.4$s. This gave us a Nyquist frequency of $17.6$mHz (Grenander, 1959, various chapters). The field of view was $38"\times20"$ and the data sequence obtained on 22.06.2004 (DS1) lasted one hour, while the data sequence obtained on 24.06.2004 (DS2) lasted $40$ minutes. The duration of the sequences was chosen to improve reliability of analysis results (Chatfield, 2003, various chapters). \par

No data were acquired specifically in support of the VTT observations, so we are unable to specify the exact state of the photospheric magnetic field during the observations. Nevertheless, the MDI instrument on SOHO was running in a mode where full disk magnetograms were obtained every $96$ minutes. Comparison of MDI with the VTT data indicates that the central VTT pointing was similar to MDI, with an error of $\pm 10"$. A G-band was used to ensure that the correct solar features were observed, i.e. no bright points in DS1 and an abundance of magnetic structures in DS2.\par

 The data set DS1 was taken in areas where no visible magnetic structuring was present. The MDI data revealed that the region scanned within the VTT field of view was in a typical quiet region, mostly unipolar (DS1, from 22.06.2004), with a moderately intense magnetic network in the data set DS2, from 24.06.2004. A filtergram of the data set DS2 shows an abundance of g-band bright structures.\par

 Figs. 1 and 2 reveal the magnetograms taken by the MDI instrument at times corresponding to the time of the observations.  It is obvious that both data sets are taken near the solar disc centre without significant difference in the inclination angle of the lines of sight.

}
\end{multicols}

{\ }

\centerline{\includegraphics[width=17cm]{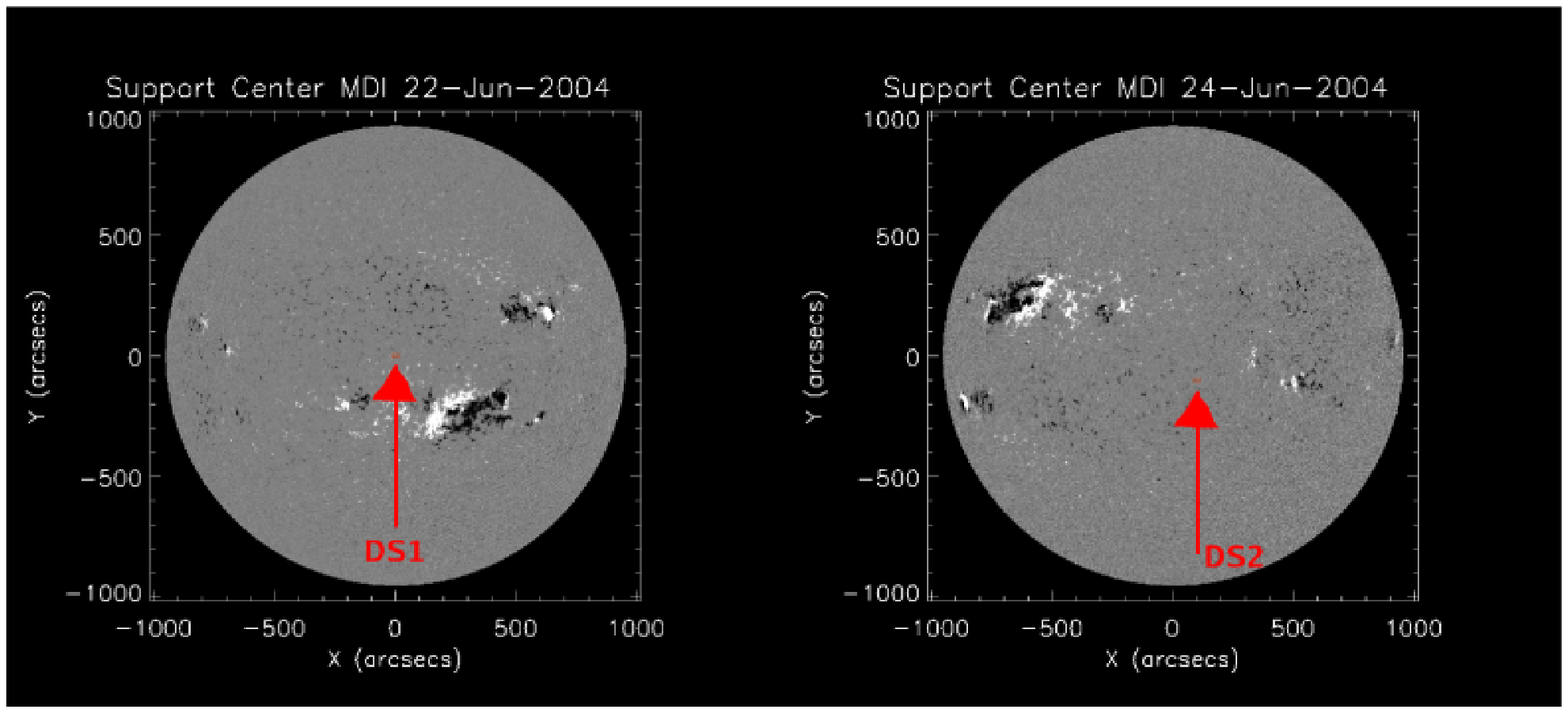}}
\figurecaption{1.}{The magnetogram of the whole solar disc from the SOHO instrument, MDI. The left panel represents the magnetogram taken on 22.06.2004 just before the observations commenced, while the right panel shows the magnetogram taken on 24.06.2004. The tips of the arrows on both panels point at the location of used field of view.}

\centerline{\includegraphics[width=17cm]{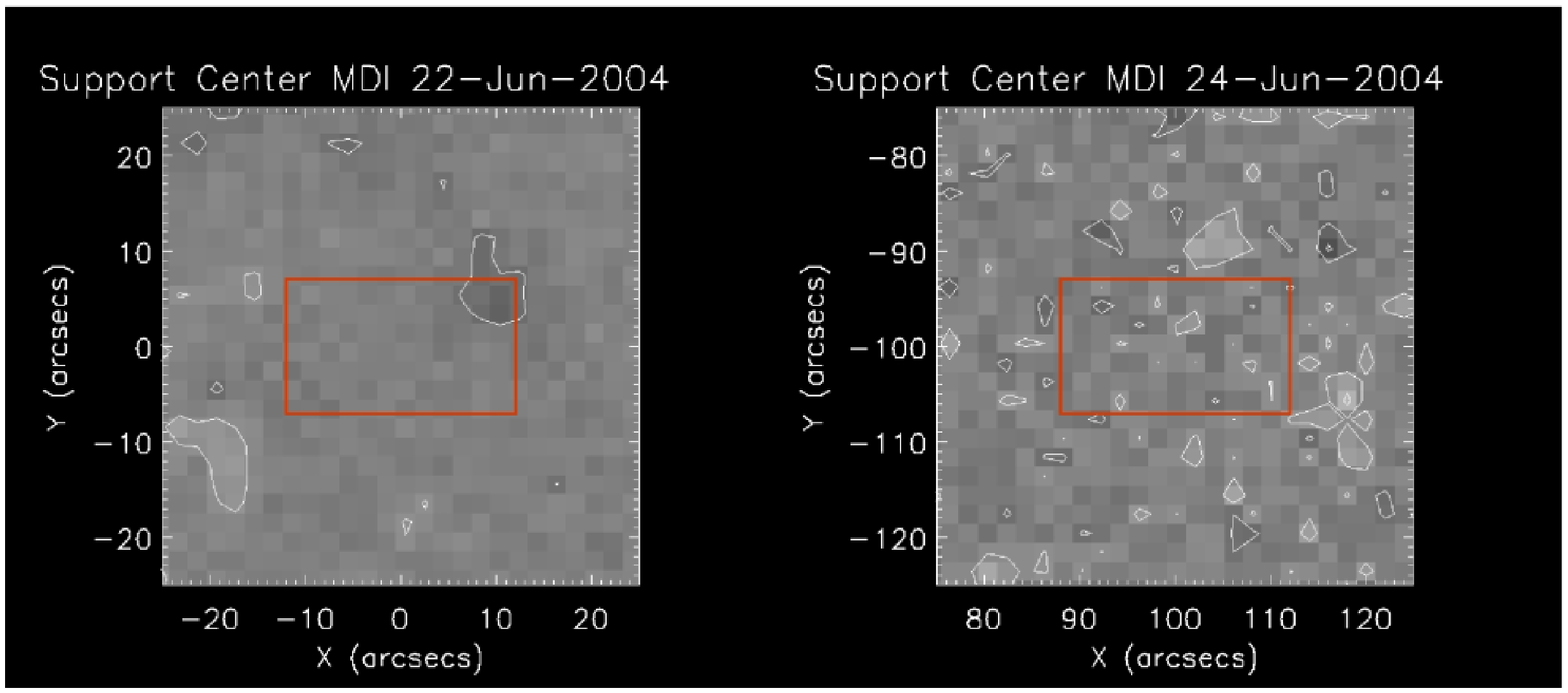}}
\figurecaption{2.}{The zoomed part of the Fig.1. The left panel represent data set DS1 and the right panel DS2. The overploted white contours are curves marking a magnetic field of 10 Gauss. The red squares represent the approximate location of the field of view.}

\begin{multicols}{2}
{

Based on these data, we conclude that the VTT time series data set DS1 corresponds to typical quiet Sun conditions, while DS2 has part of the magnetic network in the field of view (Phillips, 1992).

\section{3. DATA REDUCTION AND ANALYSIS METHODS}

The data were subjected to the reduction methods already explained in previous papers (Andic, 2007a, 2007b and Andjic, 2006). \par
As a consequence of the reduction methods and implemented subsequent coaligment of the reconstructed images the field of view and duration of the data sets was reduced. Consequently, data set DS1 has a field of view $32.3"\times14.4"$ with a duration of $52.54$min, while data set DS2 has a field of view $29.3"\times7.4"$ with a duration of $33.13$min after reduction.\par
We estimate the formation layer of the line cores (Phillips, 1992; del Toro Iniesta, 2003). We based the estimate for the formation height of the Fe {\sc i} $543.45$nm and $543.29$nm line cores on the location of optical depth unity at the central wavelength as calculated from a 3-D radiative-hydrodynamic simulation of Asplund {\it et al.} (2000).  This method is described in detail in the work of Shchukina and Trujillo Bueno (2001). The formation heights in NLTE are $258.5$km and $588.7$km for Fe {\sc i} $543.29$nm and Fe {\sc i} $543.45$nm, respectively (N. Shchukina, private correspondence).\par
The wavelet analysis used here is described in detail in previous papers (Andic, 2007a, 2007b and Andjic, 2006).  The following criteria are applied to remove spurious oscillations:

\item {*}The signal curve is tested against spurious detections of power that may be caused by Poisson noise, assuming that it is normally distributed and follows a $\chi^2$ distribution with two degrees of freedom. A confidence level of 99\%\ is calculated by multiplying the power in the background spectrum by the values of $\chi^2$ corresponding to the 99th percentile of the distribution (Torrence and Compo, 1997; Mathioudakis et al., 2003).
\item {*}The signal curve is compared with a large number (1500) of randomised time series with an identical counts distribution. By comparing the value of power found in the input signal curve with the number of times the power transform of the randomised series produced a peak of similar power, the probability of detecting non-periodic power is calculated for the peak power at each time-step.  This information is used to remove all spurious power from our results. (Banerjee et al., 2001)
\item {*}All oscillations of duration less than 1.5 cycles were excluded by comparison of the width of the peak in the wavelet power spectrum with the decorrelation time. This is undertaken to distinguish between a spike in the data and a harmonious periodic component at the equivalent Fourier frequency, thus defining the oscillation lifetime at the period of each power maximum as the interval of time from when the power supersedes 95\%\ significance until it subsequently dips below 95\%\ significance (McAteer et al., 2004). To obtain the number of cycles the lifetime is divided by the period.
\item {*}All oscillations with a power of less than 15\%\ of the maximum power in the time sequence were excluded. Wavelet analysis with the Morlet wavelet as the mother wavelet has a tendency to also detect spurious oscillations. Wavelet analysis is performed for the flat-fields, which contain noise, and the power of the oscillations detected is noted and compared with those found within the dataset. This gives us a lower limit to the power of the observed oscillations.
\item {*}All oscillations with periods of more than 788.1 s (DS1) and 496.95 s (DS2) were excluded, since they might be due to edge effects arising from the finite time span of our data.

To study the propagation characteristics of the detected waves (Andjic, 2006; Andic, 2007a and 2007b), wavelet phase coherence analysis is used (Bloomfield et al., 2004). When the power contained in the time series is studied, differences between traditional Fourier analysis and wavelet analysis are small. Fourier analysis, without timing information, has marginally different confidence levels when compared to wavelet analysis averaged over time.\par 

 However, the phase relations between the wave packets can differ in time or with a change in the local topology. These changes will be lost in Fourier analysis due to the lack of timing information. This is mostly because phase can take negative values as well, while power is always positive. Therefore, wavelet analysis is applied in this study. In the work by Bloomfield et al. (2004), a detailed description of this method is given. The equations used for the calculation of the phase difference and the phase coherence are stated in Table 1. in the work by Bloomfield et al. (2004). For pure noise this procedure yields positive coherence. Only results with a coherence above $0.6$ can be regarded as significant. \par

Phase difference and phase coherence (Davis, 2000, various chapters) are calculated using the intensity and velocity signals from both line cores.  Velocity signals were obtained from the spectral line bisector shift (Andic, 2007a, 2007b and Andjic, 2006). This makes it possible to form phase differences between velocity-velocity (V-V) signals, velocity - intensity (V-I) signals and intensity - intensity (I-I) signals. Signals are chosen from the same spatial location (i.e. same pixel in the field of view). For V-V and I-I phase difference calculations, signals are taken from the different spectral lines, while V-I phase differences are calculated from the same spectral line. This means that signals used for the V-V and I-I phase difference calculations are separated from each other by the height difference of the line cores in addition to the constant $13.5$s gap  caused by the scan time of the observed line profiles.\par

In order to compare the results with previous work, phase difference spectra are formed using the time-averaged phase differences (Fig.3). The phase difference spectra are formed so that an upward propagating wave leads to a positive phase difference. The phase differences are presented in a weighted diagram, where the weighting is applied per sample by utilising a cross-power amplitude $\sqrt{P_1P_2}$. Furthermore, they are binned into greyscale plots that are normalised to the same maximum per frequency bin (Lites and Chipman, 1979). Fig.4 shows the observed V-V phase lag between the velocity fluctuations of the Fe I $543.29$nm and Fe I at $543.45$nm lines. The coherence presented with each phase spectrum is obtained by averaging the mean time-averaged coherence (Fig.3, panel Time av. Coherence) over all processed signal curves.

\section{4. RESULTS}

 In the context of periodic phenomena phase angle is synonymous with phase. We calculate the phase using  as the reference point the oscillatory signal observed in one of the spectral lines; therefore, the phase difference is synonymous with the phase angle in this work. The term phase spectra represents distribution of the phase angle (i.e. phase difference) arranged in a progressive series according to frequency.
The calculated phase spectra, in both data sets, shows that the observed high-frequency waves do propagate. Fig.3 represents the result of the phase analysis using the wavelets on a pair of curves from data set DS2. The panels marked with LC1 Power and LC2 Power (where LC stands for light curve), represent the results of the wavelet power transforms. Lighter shaded regions correspond to an increased wave power. The contours present $95$\% confidence levels, while crosshatched areas mark the  cone of influence (COI) where edge effects can be important (Torrence and Compo, 1997; Bloomfield et al., 2006). The panel marked with Cross-wavelet Power represents the cross-wavelet power transform of both curves (in this case velocity curves), LC1 and LC2. The panel marked with Phase Difference presents full time series wavelet phase difference transforms as a function of time (abscissa) and wave frequency (ordinate). The overploted contours present $10$\% coherence exceedance levels, while the crosshatched area is COI. The panel marked with Time av. Phase Difference represents time-averaged phase differences (ordinate) from the panel marked with Phase Difference, as a function of the wave frequency(abscissa). Midpoints correspond to the mean temporal phase difference. The panel marked with Coherence is the full time series wavelet phase coherence transform. Markings are the same as for the 'Phase Difference' panel. The 'Time av. Coherence' represents time-averaged phase coherence (ordinate) of coherence as a function of wave frequency (abscissa), where the dashed line marks mean coherence.\par
The 'Phase Difference' panel shows the propagation of the high-frequency waves in different directions. Lighter shades mark upward propagation. A detailed explanation of how this plot is calculated is presented in the work by Bloomfield et al. (2004).

}
\end{multicols}

{\ }

\centerline{\includegraphics[angle=270,width=14cm]{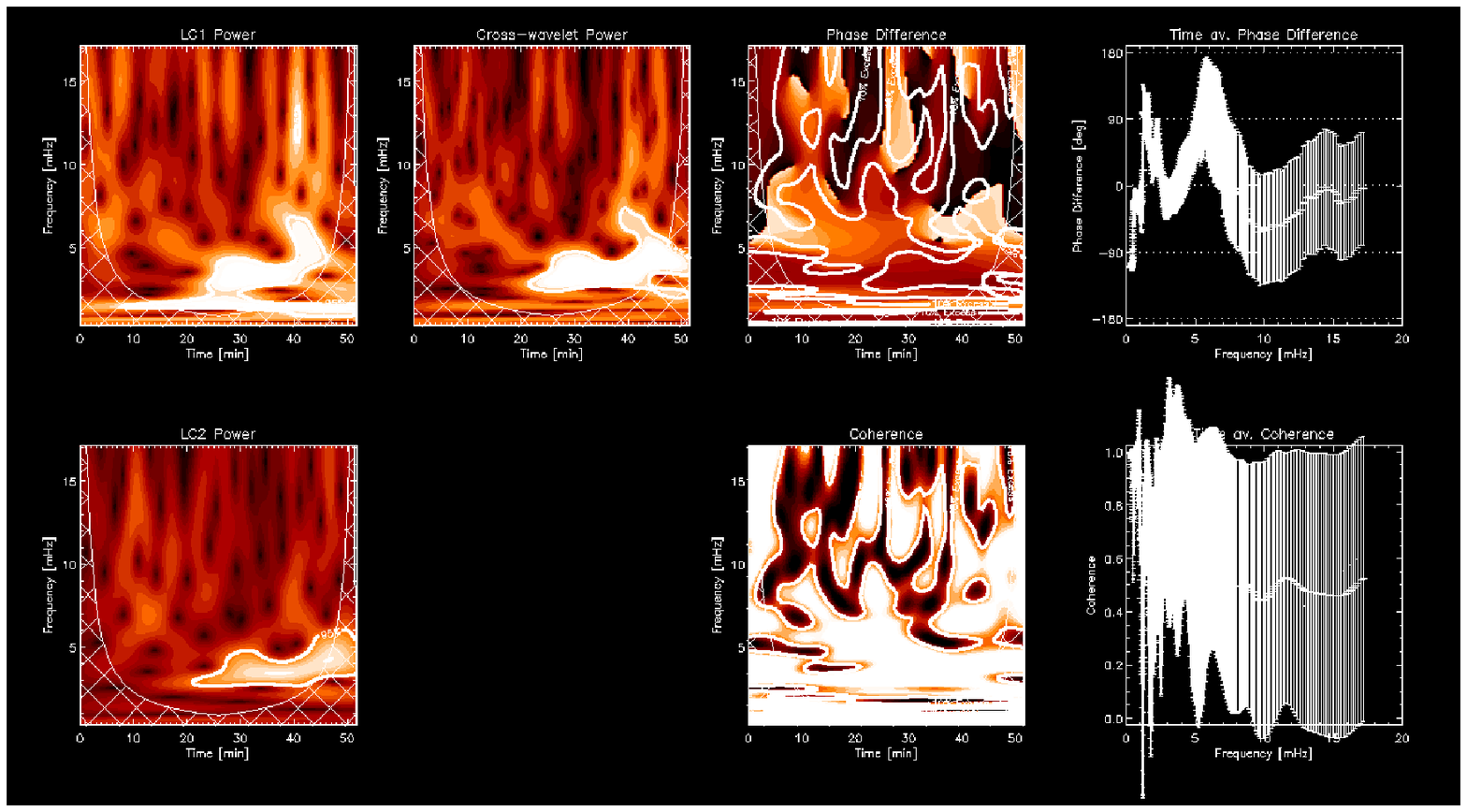}}
\figurecaption{3.}{Result of wavelet phase coherence analysis for the data set DS2. The wavelet analysis of both signal curves are shown in the panels marked with LC1 and LC2 Power. The panel marked Cross-wavelet Power shows the cross power for the both curves. The phase difference and corresponding coherence are given in the panels marked with the Phase difference and Coherence. Also shown in the last two panels are the phase difference and coherence averaged over time.}

\begin{multicols}{2}
{

 Figs.4 and 5 show that the observed direction of the propagation is upwards. Several phase spectra were calculated to obtain as much information as possible about the dynamic of the studied areas. The following results are presented separately for both data sets.

\subsection{4.1. Data set DS1}

For this dataset $46512$ signal curves were compared. The top panel of Fig.4 presents the phase spectrum calculated between velocity fluctuations (V-V) at the cores of two observed lines. The phase difference spectrum shows that the five-minute waves do not propagate. The phase difference spectrum displays upward propagation at higher frequencies up to $f \approx 13$mHz. The coherence is very small below the $1$mHz, where the disregarded frequencies are located, above that it varies between the values of $0.7$ to $1$. The phase differences for the frequencies below $1.26$mHz can be caused by the edge effects due to the finite time span. \par

The top panel on Fig.5 presents the phase spectrum calculated between intensity fluctuations (I-I) at the cores of two observed lines. The coherence distribution varies for this phase spectrum much more than for the V-V spectrum. It shows a decrease in the frequency range $4 - 8$mHz to a value close to $0.6$. After that, we can see the positive phase difference increasing. This indicates that the high-frequency waves observed in the intensity maps do travel upwards. This trend is visible even for the waves close to the Nyquist frequency, although in the frequency range after $13$mHz there is sudden drop in phase difference, even a tendency for negative phase differences as well. The coherence distribution in this range has several peaks and drops. It is significant to note that in areas where the coherence drops the positive phase difference is not dominating the spectra. 

 The black line in Fig.6 shows average velocity distribution calculated with the V-V phase spectra. The positive velocities mark upward propagation, while negative velocities mark downward propagation. On average there is no propagation for the waves in the frequency range around $5$mHz. All other frequencies show propagation. The upward propagation dominates this curve, despite a strong negative maximum near a frequency of $15$mHz. After averaging over the whole frequency range we get a propagation velocity of $26.26$m/s, which demonstrates the  domination of the upward propagation.\par
The black line in Fig.7 shows the average velocity distribution calculated with I-I phase spectra. At lower frequencies we can observe that the waves with frequencies around the cut-off frequency do propagate. The upward propagation dominates this curve. Averaged over the whole frequency range the velocity is: $19.01$m/s.\par

\centerline{\includegraphics[width=6.5cm]{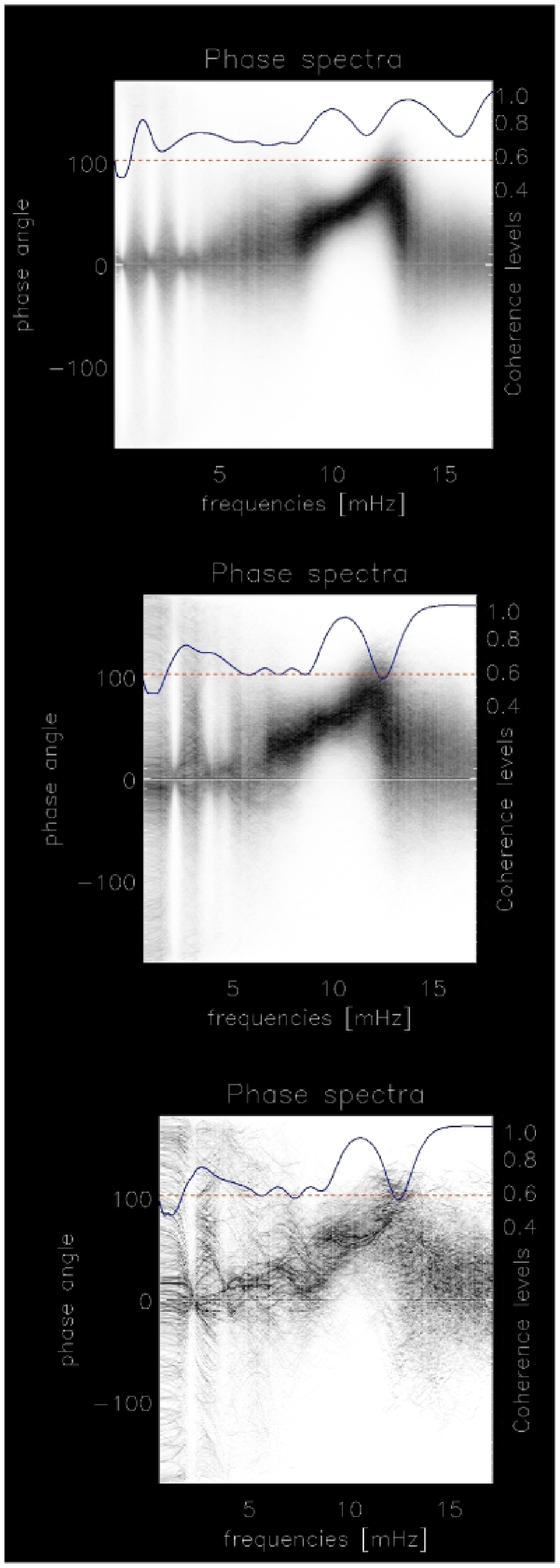}}
\figurecaption{4.}{The V-V phase difference and corresponding coherence distribution for calculated phase. The top panel represents the phase difference calculated for the data set DS1. The other two panels represent the phase difference from the DS2 data set. The bottom panel represents the phase difference for the locations where the bright points appear for at least half of the duration of a time sequence. The middle panel represents the rest of the field of view. On each panel the corresponding average coherence distribution is plotted. The dashed red lines represent coherence levels of 0.6.}
%   \label{phase22l}

\centerline{\includegraphics[width=7cm]{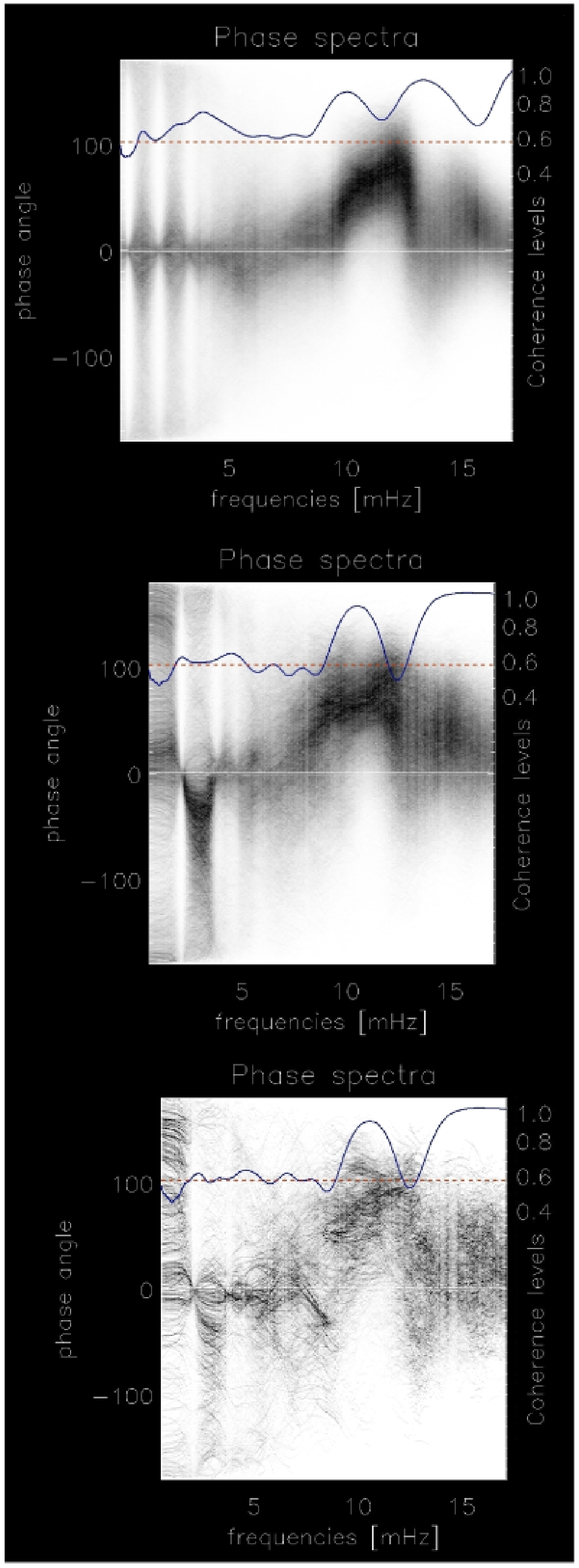}}
\figurecaption{5.}{Same as Fig.4 but the I-I phase difference is presented.}
%   \label{phase22ii}

Fig.8 shows the phase spectrum calculated between the velocity and intensity signal (V-I). The  top left panel represents results for the spectral line Fe I $543.45$nm. On average the phase spectra show that oscillations appear first in velocity curves and then in intensity curves. This trend is apparent from the concentration in frequency range of  $8 - 15$mHz. The rest of the phase spectra are ambiguous.\par

\centerline{\includegraphics[width=0.45\textwidth]{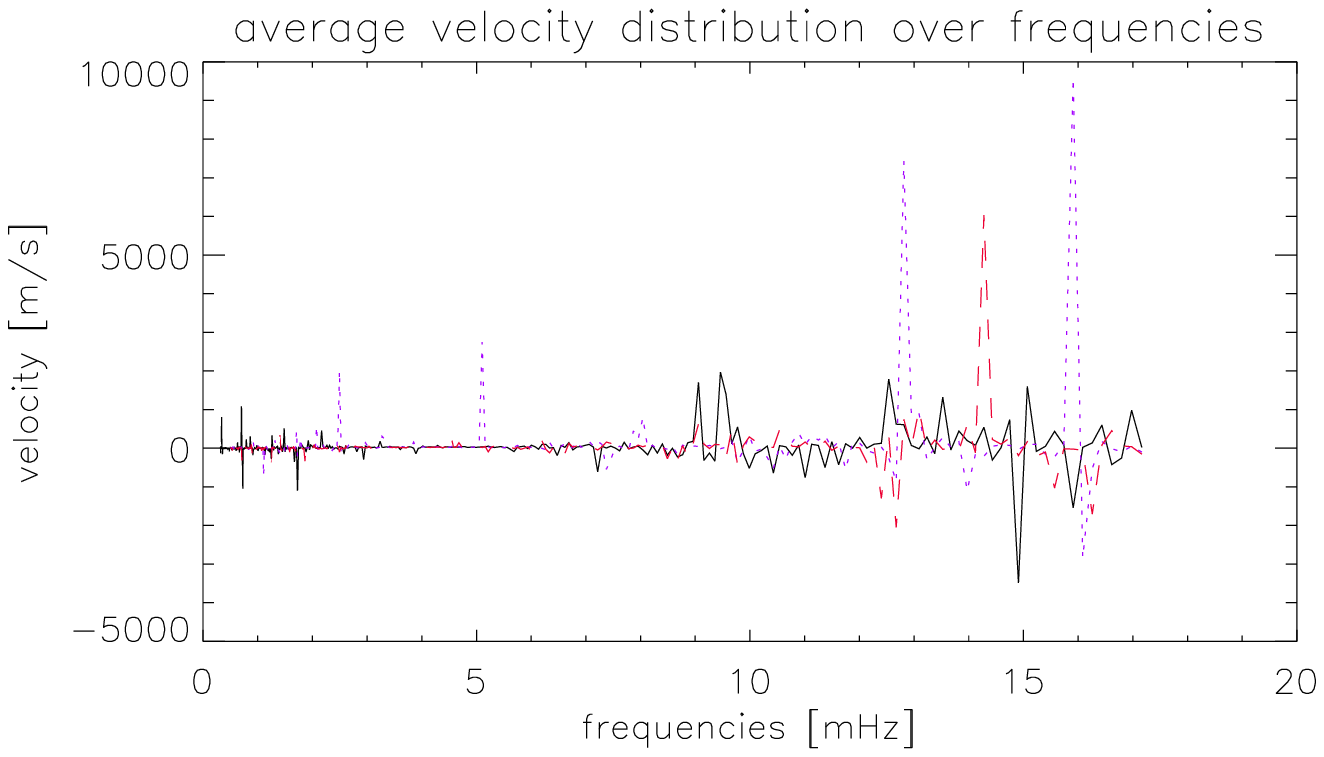}}
\figurecaption{6.}{The average velocity distribution over the analysed frequency range calculated with V-V phase difference. The black line represents distribution for the data set DS1; while the dashed red line and the dotted purple line represent the data set DS2. The dotted purple line represents results from the parts of the field of view where magnetic structures in DS2 appear for at least half of the duration of the time series, while dashed red line represents the results for the rest of the field of view.}
%   \label{vel}

The top right panel of Fig.8 presents the phase spectrum which is calculated between the velocity and intensity signal (V-I) for the spectral line Fe I $543.29$nm. The phase difference is positive at most frequencies,  indicating that signals appear first in velocity and then intensity curves. The coherence distribution is lowest for this phase spectra. The coherence distribution tends to drop in value when the intensity curves are analysed. This might be caused by the fact that intensity curves tend to have more spurious oscillations than velocity curves (Carlsson, private communication).\par

\centerline{\includegraphics[width=0.45\textwidth]{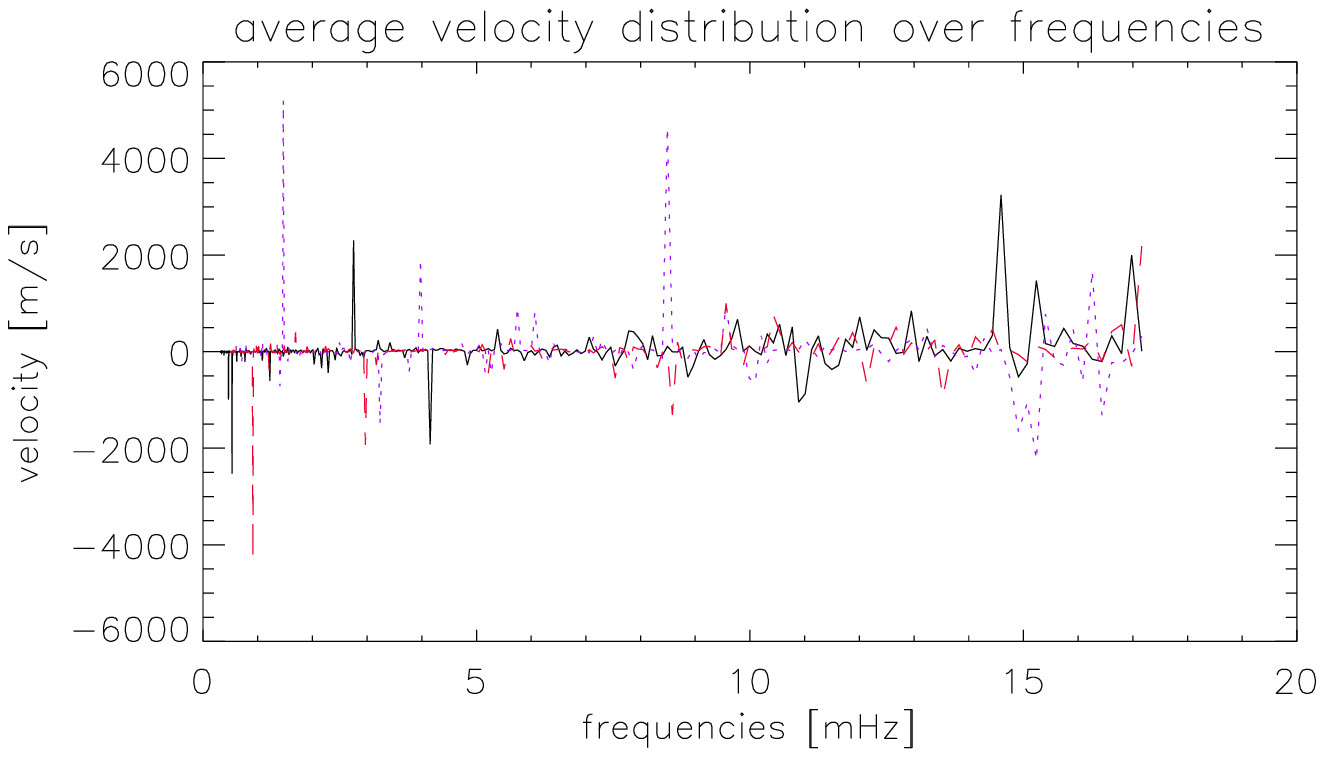}}
\figurecaption{7.}{ Same as in Fig.6 but calculated with I-I phase difference. }
%   \label{int}

\subsection {4.2. Data set DS2}

For this dataset $21682$ signal curves were compared.  The phase differences for the frequencies below $2$mHz can be caused by the edge effects due to the finite time span. Those frequencies are not taken into consideration in this analysis. Results for this data set tend to follow the general trend from the previous data set above the frequency $f \approx 7$mHz. The middle and bottom panels in Fig.4 illustrate this trend.  However, at lower frequencies the phase difference is approximately zero. Moreover the coherence distribution is relatively low, and varies much more than for the previous data set. The phase difference is calculated separately for the areas where the bright-points appear (bottom panel) and separately  for other locations (middle panel). Except the obvious lower number of curves there is no difference in behaviour. \par

The real difference can be noticed for the average velocity distribution. The dotted purple line in Fig.6 shows the average velocity distribution calculated with V-V phase spectra for the areas where the bright-points appear. In the frequency range below $6$mHz the average velocities show that there is more propagation for areas where the bright-points appear. Also, in the high-frequency range, above $12$mHz, there is a tendency for greater velocities. The dashed red line represents the average velocity distribution for the rest of the field. This line shows that the behaviour in the rest of the field is more similar to the behaviour of DS1 (solid black line). Averaged over the whole frequency range the velocity is: $74.43$m/s for areas where the bright-points appear and $23.88$m/s for the rest of the field of view.\par

An average velocity distribution calculated with I-I also shows different behaviour in the lower-frequency range for different fields of view. The view with the intense magnetic flux concentrations (dotted purple line, in Fig.7) shows different tendencies for propagation with greater velocities. The rest of the field of view (dashed red line) shows behaviour similar to DS1 (solid black line). Averaged over the whole frequency range the velocity is: $22.52$m/s for areas where the bright-points appear and $10.47$m/s for the rest of field of view. \par

The phase spectrum for the intensity waves shows similar behaviour to DS1. Where the coherence is reasonably high we can observe that the waves do travel upward (middle panel in Fig.5). The overall the coherence is much lower than for the previous data set.  The bottom panel represents results for the areas where the bright-points appear. Apart from the obvious smaller number of curves there is no difference in trend. \par

The phase spectrum V-I calculated for the spectral line Fe I $543.45$nm is presented in middle and bottom row of Fig.8. Here the coherence is very variable and lower than for the DS1.  The middle and bottom rows of Fig.8 show the phase spectra for V-I signal. The middle panels show trends that are similar to the top panels (DS1). For the spectral line Fe I $543.45$nm (left middle panel of Fig.8) it is obvious that waves appears first in velocity and then in intensity curves for the frequency range $8-13$mHz. The rest of the spectra are more ambiguous. The spectral line Fe I $543.29$nm (middle left panel of Fig.8) shows a slightly larger concentration of the phase difference in a positive range for the frequencies above $5$mHz, similar to those observed for DS1 (top left panel of Fig.8). Both show the tendency for waves to appear first in velocity and then in intensity curves.\par

V-I phase spectra, in the right bottom panels of Fig.8, are ambiguous due to the small number of curves. The tendency for the positive phase difference is clearly noticeable only above frequency $11$mHz. The left bottom panel of Fig.8 shows behaviour  similar to the left middle panel.

\section{5. DISCUSSION AND CONCLUSIONS}

The data sets used in this work were also used in previous papers (Andic, 2007a, 2007b and Andjic, 2006). The possible problems in observing high-frequency waves were discussed in detail in those papers.\par
 To observe the wave propagation one has to observe the wave at two different places in space-time. In this work, we observed the waves that leave signatures of propagation in the cores of the two different spectral lines. The signals we observed do appear first in the line core of Fe {\sc i} $543.29$nm  and then in the line core of Fe {\sc i} $543.45$nm. This should indicate wave propagation. To interpret these observations we used the model described in Asplund {\it et al.} (2000). And on the basis of this model we used the method described in detail in Shchukina and Trujillo Bueno (2001) to obtain the atmospheric heights on which observed spectral lines are formed. However, as described in several chapters of the book by del Toro Iniesta (2003) the assumption that a certain spectral line is formed at a certain atmospheric height is valid only as much in the used atmospheric model. Therefore, we have to state that the interpretation and all quoted velocity values are heavily dependent on the model and they are correct only as much as the assumptions of the model. \par
An additional problem is the assumption that the signals observed in both line cores are separated. Only under that assumption can the observed phase difference actually indicate the wave propagation. Unfortunately, this assumption is also based on the atmospheric model and the assumed independence of the line formation of the different spectral lines. Therefore, this assumption bears similar uncertainty as formation heights. \par
 In both data sets no propagation for the low-frequency waves - below $5$mHz - was observed. This is expected since most of the models (Ulmschneider, 1971a, 1971b, 2003; Stein and Leibacher, 1974; Kalkofen, 1990, 2001; Wedemeyer-B\"{o}hm et al. 2007) predict that in the lower Solar atmosphere waves with such frequencies will be evanescent, i.e. non-propagating. \par
This work concentrates only on the waves propagating in the direction parallel to the line of sight. This preference is initially chosen due to previous works on the modeling of the photospheric magnetic field. Subsurface magnetic fields are continually stretched and distorted by convective flows [Emonet \&Cattaneo 2001] resulting in the emergence of flux bundles in the photosphere. Newly emerged fields are swept towards the boundaries of granules and super-granules where they interact with pre-existing magnetic fields. In mixed-polarity regions, with mean flux density of a few Gauss, the recycling time scale is about 40 hours [Title \& Schrijver 1998]. The flux tubes are less dense than their surroundings and therefore buoyant. This buoyancy keeps the flux tubes nearly vertically oriented in the photosphere. A flux tube with a weak magnetic field in thermal and hydrostatic equilibrium with its surroundings is unstable against vertical displacements of plasma within the tube [Parker 1978]. This implies that in the quiet photosphere most of the magnetic flux tubes are vertical (i.e. parallel to the line of sight). \par
However, recent studies have shown complexity of the magnetic fields of the photosphere and chromosphere. Flux tubes, which usually channel the oscillations upward, have different propagation conditions, angle of the spread and orientation at the heights analysed here. [Ballegooijen \& Hasan 2003, Fig.6] Due to this complexity of the magnetic fields in the photosphere and chromosphere it is not clear what percentage of the oscillations propagating upward can be determined with observations along the line of sight is normal to the solar surface. To fully analyse waves propagating along the inclined flux tubes the information about the magnetic field is necessary. However, MDI does not give us information on the comparable resolution with our ground-based data. Therefore, we analysed only the waves propagating parallel to the line of sight. To analyse the rest of the waves one needs high-resolution magnetograms.

\subsection{5.1. Data set DS1}

In this data set wave propagation was observed in the frequency range from $\approx 7$ to $13$mHz. Although Krijger et al. (2001) claims that near this range there is steepening into weak shocks of acoustic waves on the way up, the resolution of the spectral line in that area makes it impossible to confirm this conclusion in our case. Since our Nyquist frequency is $17$mHz, close to this frequency we have only two points per wave and this introduces uncertainty into the calculated phase differences for those frequencies.  Also, the steepening which Krijger et al. (2001) observed starts in the range where we still observe a clear upward propagation in V-V maps (top panel of Fig.4).  There is the possibility that the drop in the phase difference after $12$mHz and subsequent reduction in the concentration of the phase differences is due to an increase in the propagation velocities of the waves. This matter will be resolved with more sensitive instruments.  The propagation is less clear in the I-I spectra (Top panel of Fig.5), where there is additional phase difference concentration around $15$mHz.\par

Moreover, top panel Fig.5 shows that around a frequency of $4$mHz there is a very slight tendency towards negative phase differences just after $5$mHz, which indicates that the observed waves are travelling downwards.  This is confirmed by the average propagation velocity distribution that shows a negative peak in the black curve in Fig.7. Both findings indicate that in this range there is a tendency for downward waves. However, the V-V phase difference in the top panel of Fig.4 does not show such a trend. This might indicate that either the sources of the velocity and intensity signals around this frequency range are not the same or there is an influence of NLTE effects in the core of the higher spectral line, Fe I $543.45$nm. \par 

The top two panels of Fig.8 show the phase spectrum calculated between the velocity and intensity signal (V-I) for the spectral lines  Fe I $543.45$nm (left top panel) and Fe I $543.29$nm (right top panel). These panels show  that the phase spectrum for the spectral lines tend to concentrate in a positive range (angle values from 40 to 100).  This trend is obvious for the frequency ranges $8-14$mHz in the top left panel of Fig.8. The top right panel does show similar tendencies for the spectral line Fe I $543.29$nm. Both panels indicate that the observed waves appear first in velocity and then in intensity curves.

}
\end{multicols}

{\ }

\centerline{\includegraphics[width=12cm]{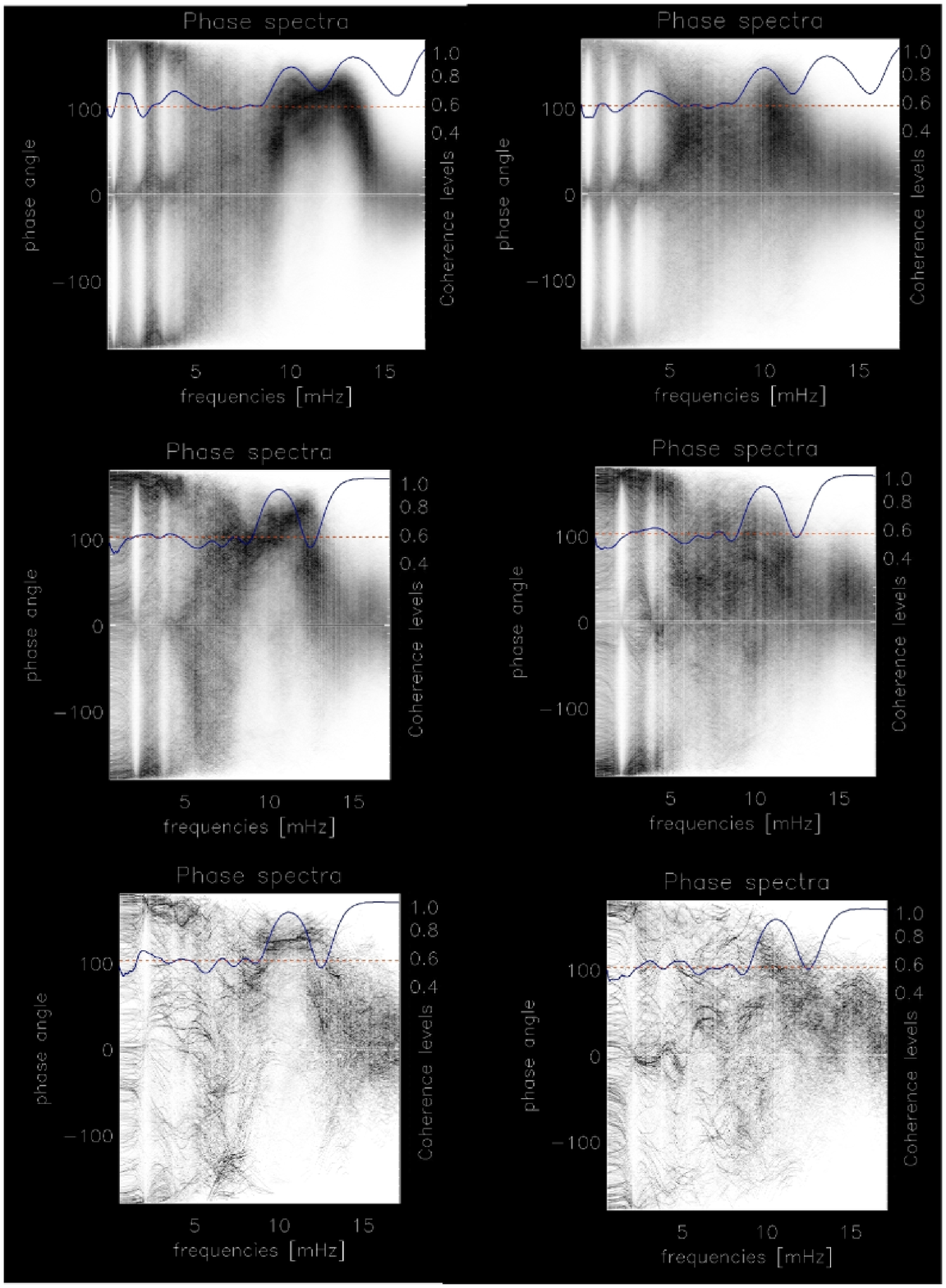}}
\figurecaption{8.}{The V-I phase difference and corresponding coherence distribution for calculated phase, as in Fig.4. Left column represents phase difference calculated for the spectral line Fe I $543.45$nm and the right one for the spectral line Fe I $543.29$nm. The top row represents results from data set DS1, while the bottom row represents the results of the reduced field of view, where magnetic structures in DS2 appear for at least half of the duration of the time series and the middle row the rest of field of view for DS2. }
%   \label{phase22vi}
\begin{multicols}{2}
{

\subsection {5.2. Data set DS2}

Results for this data set tend to follow the general trends seen in  the previous data set in the frequency range $6$-$ 13$mHz (middle panel of Fig.4).  The same trend is visible for intensity waves where the frequency range $7-12$mHz shows a clear propagation (middle panel of Fig.5). \par

Here $21682$ signal curves were compared.  We investigate whether intense magnetic flux concentrations can cause the differences noted in phase spectra. Separate phase differences are calculated for the areas where the magnetic flux concentrations appear more often than in the rest of the field of view (at least for a half of duration of the time series). Phase difference trends for those areas are similar to the rest of the field of view, as shown in the bottom panels of Figs. 4 and 5.  The only apparent difference arises from the smaller number of the analysed curves. The difference between areas with a different magnetic flux concentration is more noticeable in Figs. 7 and 6. Both diagrams show that the average propagation velocity distribution tends towards greater values for the areas where the magnetic flux concentrations appear more often (dotted purple line).\par

The middle and bottom rows of Fig.8 show the phase spectra for the V-I signal. Both show the tendency for waves to appear first in velocity and then intensity curves. The middle panels show trends that are similar to the top panels (DS1). The bottom rows show clear differences be cause of the smaller number of analysed curves. This difference is especially noticeable for the right bottom panel. Where the phase spectra trend towrds a positive phase difference is clearly noticeable only above frequency $11$mHz.  \par

Magnetic flux tubes, which usually channel the waves upward, have different propagation conditions, angles of the spread and orientation at the heights analysed here (Ballegooijen and Hasan 2003, Fig.6). This might explain the higher phase velocities registered for the waves from the area where magnetic structures are appearing for at least half of the time series duration. Work by De Pontieu et al. (2004) confirms that inclined flux tubes are the cause of the leak of p-modes energies to higher layers of the solar atmosphere. This might result in the noted positive phase velocity in the frequency ranges where one expects evanescent waves. Also, the work by Berger et al. (2004) states that the magnetic field in the plage region is concentrated in complex structures, which are not generally composed of discrete magnetic flux tubes.  This might explain why we do not observe typical behaviour for evanescent waves (Deubner and Fleck, 1989, 1990;Fleck and Deubner, 1989; Deubner et al., 1990) in our phase spectra for lower-frequency waves. \par

\subsection {5.3. Both data sets}

Krijger et al. (2001) claim that registered high-frequency waves do propagate and that phase spectra tend to lose coherence and spread for higher frequencies because of the steepening into weak shock of acoustic waves on the way up. The behaviour of the phase differences analysed here follows this pattern (Figs.4 and 5). However, the frequency ranges of the observed waves which propagate and which lose coherence differ slighty to those found by Krijger et al. (2001).  We observe propagation in the range $6 - 13$mHz for velocity signals and $8 - 13$mHz for intensity signals. The phase differences noted here are also larger than in Krijger et al. (2001). The larger phase differences might indicate detection of faster waves. On the other hand, the heights that are sampled in this work ($258.5$km and $588.7$km for Fe {\sc i} $543.29$nm and Fe {\sc i} $543.45$nm, respectively) are different to the ones sampled by Krijger et al. (2001). In the work by Krijger et al. (2001) the difference between sampled heights is approximated as $140$km while here it is $330$km. This difference in the sampled height can also allow for the larger phase angles detected here.\par

There are differences in the ranges and behaviour of the waves detected in both intensity and velocity curves  between this work and Krijger et al. (2001). Nevertheless, both phase spectra, V-V and I-I (Figs.4 and 5) show that high-frequency waves observed in this work do propagate upwards.  At frequencies around $13$mHz the peak in the phase difference is noticeable (Figs.4 and 5). This is not observed in previous work, to our knowledge. This same peak is not so clear in the middle panel of Fig.5.  In both figures, (Figs.4 and 5) is obvious that above $13$mHz it is harder to establish propagation. Similar behaviour is noted in the work by Krijger et al. (2001), but for different frequencies. Krijger et al. (2001) claims that this is due to steepening of high-frequency waves into weak shocks. Here such a statement cannot be confirmed.  Our results give three possibilities: first, steepening into shocks; second, fast waves that have angles above 360 degrees; third, the behaviour is a consequence of the low signal resolution around those frequencies.  Also, there are several possible theoretical interpretations for the noted behaviour of high-frequency waves observed here (Ulmschneider, 2003; Kalkofen, 1990, 2001; Wedemeyer-B\"{o}hm et al., 2007; Rosental et al., 2002; De Pontieu et al., 2004). Possibly the matter will be resolved in the future with instruments with higher temporal and spatial resolution.\par

Coherence distributions for each spectrum were calculated for the signal curves after the application of all conditions that were imposed to reduce spurious signals. Frequencies near the Nyquist frequency and frequencies below $1$mHz (DS1) and $2$mHz (DS2) were not cut out of the phase spectra and coherence distributions. Nevertheless, findings in those frequency ranges are not conclusive due to the unreliability of those ranges (see Sec.2).\par 

The distributions of the average velocities for the intensity and velocity signals, (Figs.6 and 7) also confirm that there is a dominating trend for upward propagation. Stating finite numbers for phase velocities has to be done with caution. Except the influences of the solar conditions at the formation heights themselves (Asplund et al., 2000) there is also an influence caused by post-focus instrumentation (Bendlin et al., 1992) and effects of transmission of the telescope (Berkefeld, Soltau and von der Luehe, 2003). Therefore, all those factors have to be taken into account when discussing and analysing the results of phase analysis. The phase angles show that there is propagation, but values of the propagation velocities are highly dependant on the assumed values for the height and therefore suffer all of the  problems connected with such assumptions. However, the values of the observed phase velocities are greater in the areas where magnetic structures appear for longer than half of the time series' duration. The fact that propagation velocities are greater in the areas with significant magnetic concentrations is in agreement with the work by Rosental et al. (2002). These authors claim that, in the presence of the magnetic field concentration, the propagation velocities of high-frequency waves will be greater. Detection of the propagation in the low-frequency range, for the same data sample from DS2, is in agreement with the work by Jefferies  et al. (2006) and De Pontieu et al. (2004). Both works claim that the presence of the magnetic field allows the low-frequency waves to propagate.\par
 However, since we did not have high resolution magnetograms for the data sets, the exact influence of the magnetic field cannot be stated. Although there are indications that magnetic fields can be responsible for such behaviour, there is a possibility that such behaviour is a consequence of the stochastic behaviour of the wave sources (Houdek et al. 1999). This question will be resolved with the high resolution magnetograms.\par
Also, there is a significant difference in the diagrams of the intensity and velocity signals. This difference can be explained with the properties of the Radiative Transfer Equation (summary explanation with various references can be found in several chapters of the book by del Toro Iniesta 2003). In short, from a theoretical standpoint, velocity signals shows waves clearer than intensity signals. However, this does not mean that results from intensity are less reliable and can be ignored. The observed behaviour of both signals can help in improving the future atmospheric models that will deepen our understanding of the solar atmosphere.

Fig.8 shows the phase spectrum calculated between velocity and intensity signals (V-I). All panels, except the  bottom right one, show a clear tendency for waves to appear first in velocity and then after in the intensity signal. The top right panel (spectral line Fe I $543.29$nm, DS1) both middle panels (spectral lines  Fe I $543.45$nm (left ) and Fe I $543.29$nm (right),DS2) and the left bottom panel (spectral line Fe I $543.29$nm, DS2) show a tendency for positive phase angles. The trend for angles of $\approx 90^{\circ}$ for the low-frequency range $3-6$mHz is stated as typical for evanescent waves in the work of Deubner et al. (1990). In Fig.8 the middle left panel is the only one where this tendency can be seen for frequencies around $6$mHz. Therefore we cannot certify this as typical behaviour for evanescent waves.\par
Deubner and Fleck (1990) offers V-I phase diagrams for chromospheric lines which are significantly different to the V-I diagrams found here, indicating that the waves observed here do not exhibit the behaviour noted for the chromospheric waves. This difference might be caused by the heights sampled here. Although Deubner and Fleck (1989) sample photospheric and chromospheric heights, the frequency ranges are difficult to compare with our results. As stated in section 2, frequency ranges below $3$mHz might be influenced by edge effects arising from the finite time span of our data. This gives us only frequencies  $3-5$mHz to compare. Unfortunately there is no visible agreement. Clear tendencies noticed by Deubner and Fleck (1989), Fleck and Deubner (1989) and Deubner et al. (1990) that indicate evanescent waves are not detectable in our data. It is possible that we do observe propagation of such waves, as predicted by  De Pontieu et al. (2004) and commented on by Jefferies  et al. (2006) as the possible cause of chromospheric heating. Comparing our results with works by Deubner and Fleck (1989, 1990), Fleck and Deubner (1989) and Deubner et al. (1990) can lead to the conclusion that we observe several different kinds of waves simultaneously, which are expected in light of recently developed multidimensional models and theoretical explanations of the solar atmosphere (Rosental et al., 2002; De Pontieu et al., 2004; Wedemeyer et al., 2004; Wedemeyer-B\"{o}hm et al., 2007). \par

In Figs.4, 5 and 8 one can notice two clear streaks around frequencies below $3$mHz. As stated in section 2 those ranges might be influenced by edge effects arising from the finite time span of our data, therefore those ranges were excluded from discussion and analysis here.\par

The observed waves propagate upward in the frequency range $8$-$ 13$mHz.  We also noted indications that in the lower-frequency range, below $8$mHz, there is an indication of the propagation in areas where the magnetic flux concentrations appear.

% Acknowledgements

\acknowledgements{I am grateful to the Science and Technology Facilities Council for financial support. I wish, especially, to thank Dr. N. Shchukina for calculating the formation heights for the lines used.
Also, I wish to thank M. Mathioudakis and E. Wiehr for stimulating discussions. For help with the observations I wish to thank J.K. Hirzberger and K.G.Puschman. I wish to thank F. Kneer for insisting on the importance of the research of these oscillations. Also I wish to thank A.M. Broomhall for advices on language.}

% References

\references

Andjic,A. :2006, \journal{ Serb.Astron.J.} \vol{ 172}, 27.

Andic,A.,  \journal{Solar Phys.},  \vol{242}, 9.

Andic,A.,  \journal{Solar Phys.},  \vol{243}, 131.

Asplund,M., Nordlund,A., Tramperdach,R., Allende Prieto,C., Stein,R.F. :2000,  \journal{Astron. Astrophys.}  \vol{359},729.

Bendlin,C., Volkmer,R., Kneer,F. :1992,  \journal{Astron. Astrophys.}  \vol{257}, 817

Banerjee, D., O'Shea, E., Doyle, J. G., Goossens, M., 2001, \journal{Astron. Astrophys.}  \vol{371}, 1137

Berger, T.E, Rouppe van der Voort, L.H.M., L\"ofdahl,M.G., Carlsson,M., Fossum,A., Hansteen, V.H., Marthinussen,E., Title,A., Scharmer,G.: 2004,  \journal{Astron. Astrophys.} \vol{428},613.

Berkefeld,Th., Soltau,D. von der L�he,O. :2003,  \journal{Proceedings of the SPIE}  \vol{4839}, 544

Bloomfield,D.S., McAteer,R.T.J., Lites,B.W., Judge,P.G., Mathioudakis,M.,  Keenan,F.P. :2004  \journal{Astrophys. J.}  \vol{617}, 623.

Bloomfield,D.S., McAteer,R.T.J., Mathioudakis,M., Keenan,F.P. :2006  \journal{Astrophys. J.}  \vol{652}, 812.

Chatfield, C.,  \journal{ The Analysis of Time Series: An Introduction (Chapman \& Hall Texts in Statistical Science Series)}, Chapman \& Hall/CRC; 6th Ed edition, p.352

Davis, J.L. :2000,  \journal{Mathematics of Wave Propagation}, Princeton University Press, p.416

del Toro Iniesta, J.C. :2003, \journal{Introduction to Spectropolarimetry}, Cambridge University Press, p.227

Deubner,F.-L., Fleck,B., Marmolino, C., Severino,G.: 1990,  \journal{Astron. Astrophys.} \vol{236}, 509.

Deubner,F.-L., Fleck,B.: 1990,  \journal{Astron. Astrophys.} \vol{228}, 506.

Deubner,F.-L., Fleck,B.: 1989,  \journal{Astron. Astrophys.} \vol{213}, 423.

De Pontieu, B., Erd\'elyi, R., James, S.P.:2004  \journal{Nature} \vol{430},536

Emonet, T., Cattaneo, F.:2001, \journal{Astrophys. J.} \vol{560}, L197

Fleck,B., Deubner,F.-L.: 1989,  \journal{Astron. Astrophys.} \vol{224}, 245.

Fossum,A., Carlsson, M. :2006,  \journal{Astrophys. J.}  \vol{646},579.

Grenander, U. :1959,  \journal{Probability and statistics: The Harald Cramer volume (Unknown Binding)}, Almqvist \& Wiksell; Wiley; Chapman \& Hall, p.434

Ghosh, S.N.: 2002  \journal{Electromagnetic Theory and Wave Propagation}, CRC Press Inc; 2nd Ed edition, p.259

Houndek, G., Balmforth, N.J., Christensen-Dalsgaard, J., Gough, D.O. :1999, \journal{Astron. Astrophys.} \vol{351},582

Jefferies, S.M., McIntosh, S.W., Amstrong, J.D., Bogdan, T.J., Cacciani, A., Fleck, B. :2006,  \journal{Astrophys. J.}  \vol{648}, L151.

Kalkofen,W.: 1990,In: Priest, E.R., Krishan, V. (eds.), \journal{Basic Plasma Processes on the Sun},{\it IAU Symp.}  \vol{142}, 197.

Kalkofen,W.: 2001,  \journal{Astrophys. J.}  \vol{557}, 376.

Krijger,J.M., Rutten, R.J., Lites, B.W, Straus, Th., Shine, R.A., Tarbell, T.D.: 2001,  \journal{Astron. Astrophys.} \vol{379}, 1052.

Lites,B.W., Chipman, E.G.:1979, \journal{Astrophys. J.}  \vol{231}, 570

Mathioudakis, M., Seiradakis, J.H., Williams, D.R., Avgoloupis, S., Bloomfield, D.S., McAteer, R.T.J., 2003,  \journal{Astron. Astrophys.} \vol{403}, 1101 

McAteer, R. T. J., Gallagher, P. T., Bloomfield, D.S., Williams, D. R., Mathioudakis, M., Keenan, F.P., 2004,  \journal{Astrophys. J.}  \vol{602}, 436

Parker, E.N., \journal{Astrophys. J.} \vol{221}, 368

Phillips, K.J.H.: 1992,  \journal{Guide to the Sun}, Cambridge University Press, p.386

Rosenthal,C.S., Bogdan, T.J., Carlsson, M., Dorch,S.B.F., Hansteen,V., McIntosh,S.W., McMurry, A., Nordlund,A., Stein,R.F. :2002, \journal{Astrophys. J.}  \vol{564},508.

Shchukina,N.G., Trujillo Bueno,J.: 2001,  \journal{Astrophys. J.}  \vol{550}, 970.

Stein,R.F., Leibacher, J.: 1974,  \journal{Astron. Astrophys.} \vol{12}, 407

Thio, T.: 2006,  \journal{American Scientist}  \vol{94},40

Title, A.M., Schrijver, C.J :1998, \journal{in ASP Conf. Ser.} \vol{ 154}, The Tenth Cambridge Workshop on Cool Stars, Stellar Systems and the Sun, ed. R. A. Donahue, J.A. Bookbinder (San Franciso:APS), 345

Torrence,C., Compo,G.P.: 1998,  \journal{Bull. Amer. Meteor. Soc.}  \vol{79}, 61.

Ulmschneider,P.: 1971b  \journal{Astron. Astrophys.} \vol{14}, 275.

Ulmschneider,P.: 1971a,  \journal{Astron. Astrophys.} \vol{12}, 297.

Ulmschneider,P., Review: the physics of chromosphere and corona. In: Lectures in solar Physic, H.M. Antia, Springer \& Verlag, 2003

van Ballegooijen, A.A, Hasan, S.S.:2003, Preparing for ATST ASP Conference Series,  Vol. 286

Wang, Z., Urlich, R.K., Coroniti, F.V.: 1995,  \journal{Astrophys. J.}  \vol{444},879.

Wedemeyer, S., Freytag, B., Steffen, M., Ludwig, H.-G., Holweger, H.:2004,  \journal{Astron. Astrophys.}  \vol{414},1121

Wedemeyer-B\"ohm, S., Steiner, O., Bruls, J., Rammacher, W.:2007  \journal{arXiv:astro-ph} \vol{0612627v1}

\endreferences

}
\end{multicols}

\vfill\eject

{\ }

% Serbian abstract

% Title

\naslov{PROPAGACIJA VISOKOFREKVENTNIH TALASA U TIHOJ SUNCEVOJ ATMOSFERI}

% Authors

\authors{A. Andi{\' c}$^{1,2}$}

\vskip3mm

% Address

\address{$^1$HiROS,School of Physics and Astronomy, College of Engineering and Physical Sciences, The University of Birmingham, Edgbaston, Birmingham, B15~2TT, UK}

\address{$^2$ Astrophysics Research Centre, School of Mathematics and Physics, Queen's University, Belfast, BT7~1NN, UK}

\vskip.7cm

% UDC

\centerline{UDK \udc}

% Papertype

\centerline{\rit }

\vskip.7cm

\begin{multicols}{2}
{

% Abstract

\rrm Visokofrekventni talasi  ($5$mHz - $20$mHz) su predlozeni kao izvor grijanja tihe sunceve atmosfere. Dinamika tih, predhodno detektovanih, talasa je ovde prestavljena. Sekvence slika su dobivene koristenjem Fabri-Pero spektrometar lociran U Njemackom Vakuumskom-Toranj Teleskopu (VTT) na Observatorio del Teide, Izana, Tenerife. Podaci su potom redukovani spekle metodom i analizirani vavletima. Da se utvrdi da li posmatrani talasi propagiraju provedena je analiza fazne razlike bazirana na vavletima. Zabiljezeli smo propagaciju talasa u frekventnom opsegu od:$10$mHz to $13$mHz. Takodje, zabiljezena je propagacija talasa u frekventnim opsezima gdje ti talasi ne propagiraju, na lokacijama gdje su bile prisutne magnetne strukture. 

}
\end{multicols}

\end{document}